\documentclass[amsmath,11pt]{article}

\usepackage{amssymb}

\date{\empty}

\begin{document}

\title{\bf The deceleration parameter in `tilted' Friedmann universes: Newtonian vs relativistic treatment}

\author{C.G. Tsagas${}^{1,2}$, M.I. Kadiltzoglou${}^1$ and K. Asvesta${}^1$\\ ${}^1${\small Section of Astrophysics, Astronomy and Mechanics, Department of Physics}\\ {\small Aristotle University of Thessaloniki, Thessaloniki 54124, Greece}\\ ${}^2${\small Clare Hall, University of Cambridge, Herschel Road, Cambridge CB3 9AL, UK.}}

\maketitle

\begin{abstract}
Although bulk peculiar motions are commonplace in the universe, most theoretical studies either bypass them, or take the viewpoint of the idealised Hubble-flow observers. As a result, the role of these peculiar flows remains largely unaccounted for, despite the fact that relative-motion effects have led to the misinterpretation of the observations in a number of occasions. Here, we examine the implications of large-scale peculiar flows for the interpretation of the deceleration parameter. We compare, in particular, the deceleration parameters measured by the Hubble-flow observers and by their bulk-flow counterparts. In so doing, we use Newtonian theory and general relativity and employ closely analogous theoretical tools, which allows for the direct and transparent comparison of the two studies. We find that the Newtonian relative-motion effects are generally too weak to make a difference between the two measurements. In relativity, however, the deceleration parameters measured in the two frames differ considerably, even at the linear level. This could deceive the unsuspecting observers to a potentially serious misinterpretation of the universe's global kinematic status. We also discuss the implications and the observational viability of the relativistic study.
\end{abstract}

\section{Introduction}\label{sI}
The Milky Way, as well as every other typical galaxy in the universe, has its own peculiar motion and does not follow the smooth universal expansion, also known as the Hubble flow~\cite{Pe1}-\cite{CoLu}. More specifically, our Local Group of galaxies moves at a speed of more than 600~km/sec relative to the Hubble expansion. Such large-scale peculiar motions, which are commonly referred to as ``bulk flows'', have been repeatedly verified by numerous surveys (see~\cite{Aetal} and references therein). Nevertheless, most of the theoretical cosmological studies bypass these bulk peculiar flows and nearly all of them take place in the coordinate system of the ``fictitious'' Hubble-flow observers. As a result, the full effect of our peculiar motion is still largely unaccounted for, despite the fact that the history of astronomy is rife with examples where relative-motion effects have led to a gross misinterpretation of reality.

To study peculiar motions in cosmology, as well as in general, one needs to introduce two (at least) families of observers. One family will determine the reference frame, relative to which peculiar motions can be defined and measured. In cosmology, the universal expansion provides such a preferred coordinate system. This is the idealised frame of the smooth Hubble flow, where the dipole of the Cosmic Microwave Background (CMB) vanishes.\footnote{Hereafter, we will use the terms Hubble frame and CMB frame interchangeably.} The second group of observers, on the other hand, are the ``real'' ones, namely those living in typical galaxies (like our Milky Way) that move relative to the Hubble expansion. In relativity, cosmological models equipped with two sets of relatively moving observers are usually termed ``tilted'' (e.g.~see~\cite{CE,HUW}). In these models, both the properties of the matter sources and the kinematics, as seen by the tilted observers, generally differ due to relative-motion effects. For instance, the reader is referred to~\cite{HDPI} for a thorough discussion of tilted Lemaitre-Tolman-Bondi spacetimes. In a more recent example, \cite{NS} have studied peculiar velocities in Szekeres-II regions matched to a $\Lambda$CDM background, in an attempt to alleviate the familiar ``$H_0$-tension''.

Tilted cosmologies have also been employed in a series of articles to investigate the effects of large-scale bulk peculiar motions on the deceleration parameter~\cite{T1}-\cite{T3}, while more recently they were also used to study the evolution of the peculiar velocity-field itself~\cite{TT,FT}. Initially, the motivation was to investigate whether relative-motion effects could alter the kinematics of the tilted observers and thus lead them to a misinterpretation of the cosmological data. This line of research was pursued and extended in the more recent studies as well, where the energy-flux contribution to the relativistic gravitational field was identified as the physical reason separating the relativistic from the Newtonian studies of peculiar velocities and their implications. That work also demonstrated how the relativistic treatment could drastically change the standard (Newtonian) picture. With these in mind, we provide here a combined Newtonian and relativistic analysis of the relative-motion effects on the deceleration parameter. In so doing, we employ the same mathematical tools, namely the relativistic 1+3 covariant formalism and its Newtonian version (see~\cite{Eh}-\cite{EMM} and~\cite{El2,El3} respectively). This facilitates the direct and transparent comparison between the two treatments, which in turn allows us to identify their differences and to reveal the physical reasons behind them.

Starting with the Newtonian study and adopting a nearly Friedmann-Robertson-Walker (FRW) universe, we find that the linear relative-motion effects have essentially no impact on the deceleration parameter, as measured in the aforementioned two frames. Put another way, the deceleration parameter measured by the fictitious Hubble-flow observers, which coincides with the deceleration parameter of the universe itself, is essentially identical to the one measured by the real observers in their local bulk-flow frame. In relativity, however, the differences between the two measurements can be quite substantial, even at the linear level. More specifically, the relativistic effects can make the deceleration parameter measured in the bulk-flow frame to appear (locally) negative, in a universe that is still (globally) decelerating. This theoretical claim, which was originally made in~\cite{T1}-\cite{TK} (see also~\cite{T3}), has recently received independent support~\cite{H}. Here, we confirm these claims and in the process refine, extend and further explain the recent study of~\cite{T3}. Finally, by comparing the relativistic to the Newtonian treatments, we are also able to identify the physical reasons responsible for their aforementioned disagreement.

Our work shows that, at the linear level, the Newtonian and the relativistic studies proceed in close parallel up to a certain point. Until then, the two approaches are indistinguishable for all practical purposes. The relativistic treatment starts to diverge when the role of the gravitational field is taken into account. In Newtonian theory, the latter propagates via the associated potential, with the matter contribution coming through the Poisson equation. There is no potential in relativity. There, Poisson's formula is replaced by the Einstein field equations and the matter input to the gravitational field comes into play through the energy-momentum tensor. When peculiar motions are accounted for, the local stress-energy tensor (and therefore the associated gravitational field) contains an additional energy-flux input due to the moving matter. In a sense, one might say that the bulk flow gravitates~\cite{TT,FT}. This flux contribution feeds into the Einstein field equations, then into the energy and the momentum conservation laws and eventually emerges into the formulae governing the linear evolution of the peculiar-velocity field. No Newtonian study can naturally reproduce such effects, which explains why the two theories arrive at considerably different results and conclusions. Overall, the disagreement between the Newtonian and the relativistic results identified in this work, is traced back to the fundamentally different way the two theories treat the gravitational field and its sources.

We start our presentation in \S~\ref{sCF} with a brief outline of the covariant approach to fluid dynamics, both in Newtonian theory and in general relativity. In the next section, we apply the formalism to models equipped with two families of relatively moving observers. These are related by the Galilean transformation in the Newtonian treatment and by the Lorentz boost in the relativistic. Section 4 provides the linear relations between the key kinematic variables measured by the aforementioned two groups of observers. The Newtonian relative-motion effects on the deceleration parameter are presented in \S~\ref{sNEtq} and those of the relativistic analysis in \S~\ref{sREtq}. There, we also compare the two treatments and identify the reasons behind their disagreement.

\section{Covariant formalism}\label{sCF}
The covariant approach to fluid dynamics dates back to the 1950s~\cite{HS,R}. The formalism offers a Lagrangian treatment, which was first applied to Newtonian theory, before extended to relativity and cosmology~\cite{Eh,El1}. Here, we will go through the basics of the two formalisms, referring the reader to~\cite{TCM,EMM} for extended reviews and further discussion.

\subsection{Newtonian covariant approach}\label{ssNCA}
Consider a family of observers in a 3-dimensional Euclidean space and assume that $h_{\alpha\beta}$ is the associated metric tensor, relative to a coordinate system $\{x^{\alpha}\}$. Then, $h_{\alpha}{}^{\alpha}=3$ by construction and $v^2=h_{\alpha\beta}v^{\alpha}v^{\beta}$ for any vector field $v_{\alpha}$. In a Cartesian frame, we have $h_{\alpha\beta}= \delta_{\alpha\beta}$, where $\delta_{\alpha\beta}$ is the Kronecker delta. Otherwise, $h_{\alpha\beta}\neq\delta_{\alpha\beta}$ and both $h_{\alpha\beta}$ and $h^{\alpha\beta}$ (with $h_{\alpha\mu}h^{\mu\beta}=\delta_{\alpha}{}^{\beta}$) are needed to raise and lower indices, as well as to compensate for the ``curvature'' of the coordinate system~\cite{El1,El3}. Here, we will keep using upper and lower indices, even when dealing with Cartesian vectors and tensors, to facilitate the comparison between the Newtonian and the relativistic equations. One can distinguish the Newtonian formulae from the use of ordinary partial (rather than covariant) derivatives and from the use of Greek (instead of Latin) indices.

Assuming an expanding Newtonian universe, we introduce the vector field ($u_{\alpha}$) to monitor the motion of the aforementioned observers. Relative to the $u_{\alpha}$-field, the time derivative of a general (tensorial) quantity ($T$) is the convective derivative $\dot{T}=\partial_tT+u^{\alpha}\partial_{\alpha}T$. The (inertial) acceleration of the matter, for example, is given by $\dot{u}_{\alpha}=\partial_tu_{\alpha}+ u^{\beta}\partial_{\beta}u_{\alpha}$. Further kinematic information is obtained by splitting the gradient of the velocity as~\cite{El1,El3}
\begin{equation}
\partial_{\beta}u_{\alpha}= {1\over3}\,\Theta h_{\alpha\beta}+ \sigma_{\alpha\beta}+ \omega_{\alpha\beta}\,,  \label{pbua}
\end{equation}
with $\Theta=\partial^{\alpha}u_{\alpha}$ being the volume scalar, $\sigma_{\alpha\beta}=\partial_{\langle\beta}u_{\alpha\rangle}$ the shear tensor and $\omega_{\alpha\beta}=\partial_{[\beta}u_{\alpha]}$ the vorticity tensor.\footnote{Throughout this manuscript Greek indices run between 1 and 3, while their Latin counterparts take values from 0 to 3. Also, round brackets imply symmetrisation, square antisymmetrisation and angled indicate the symmetric, trace-free component of second-rank tensors (e.g.~$\sigma_{\alpha\beta}= \partial_{(\beta}u_{\alpha)} -(\partial^{\mu}u_{\mu}/3)h_{\alpha\beta}$).} The former implies  expansion/contraction, when positive/negative, while nonzero shear and vorticity ensure kinematic anisotropies and rotation respectively. Note that the volume scalar also defines the cosmological scale factor ($a$), by means of $\dot{a}/a=\Theta/3$. Then, in a Friedmann model, the volume scalar and the Hubble parameter are related by $\Theta=3H$.

In Newtonian gravity, the potential ($\Phi$) couples to the density of the matter ($\rho$) via the Poisson equation $\partial^2\Phi= \kappa\rho/2$, where $\kappa=8\pi G$. Also, the gravitational acceleration is given by the gradient of the potential, which added to the inertial acceleration gives~\cite{El1,El3}
\begin{equation}
A_{\alpha}= \dot{u}_{\alpha}+ \partial_{\alpha}\Phi\,.  \label{Aa}
\end{equation}
Written this way, the vector $A_{\alpha}$ vanishes when matter moves under inertia or/and gravity alone (see Eq.~(\ref{Euler}) below) and therefore corresponds to the relativistic 4-acceleration (see \S~\ref{ssR1+3CA} next). On using the above, Euler's formula reads
\begin{equation}
\rho A_{\alpha}= -\partial_{\alpha}p- \partial^{\beta}\pi_{\alpha\beta}\,,  \label{Euler}
\end{equation}
where $p$ and $\pi_{\alpha\beta}$ (with $\pi_{\alpha\beta}= \pi_{\beta\alpha}$ and $\pi_{\alpha}{}^{\alpha}=0$) are respectively the isotropic pressure and the viscosity of the fluid~\cite{El1,El3}. On the other hand, the Newtonian continuity equation is
\begin{equation}
\dot{\rho}= -\Theta\rho\,.  \label{cont}
\end{equation}
Note that, throughout this study, we will assume a barotropic cosmic medium with an equation of state of the form $p=p(\rho)$.

\subsection{Relativistic 1+3 covariant approach}\label{ssR1+3CA}
In relativity, observers move along timelike worldlines tangent to the 4-velocity vector $u_a$, which is normalised so that $u_au^a=-1$. The 4-velocity field also defines the tensor $h_{ab}=g_{ab}+u_au_b$ (with $g_{ab}$ being the spacetime metric) which is symmetric and projects orthogonal to $u_a$ (since $h_{ab}=h_{(ab)}$ and $h_{ab}u^b=0$ by construction). The projector also satisfies the constraints $h_a{}^a=3$ and $h_{ac}h^c{}_b=h_{ab}$, while it acts as the metric of the observers' 3-dimensional rest-space in the absence of rotation (e.g.~see~\cite{TCM,EMM}).

The $u_a$ and $h_{ab}$ fields introduce an 1+3 splitting of the spacetime into time and 3-D space, along and orthogonal to the 4-velocity vector. Therefore, the associated temporal and spatial derivative operators are respectively given by
\begin{equation}
{}^{\cdot}= u^a\nabla_a \hspace{15mm} {\rm and} \hspace{15mm} {\rm D}_a= h_a{}^b\nabla_b\,,  \label{1+3ops}
\end{equation}
with $\nabla_a$ representing the 4-D covariant derivative. Using the above, the gradient of the 4-velocity field decomposes into the irreducible kinematic components of the observers' motion as follows
\begin{equation}
\nabla_bu_a= {1\over3}\,\Theta h_{ab}+ \sigma_{ab}+ \omega_{ab}- A_au_b\,,  \label{Nbua}
\end{equation}
where $A_a=u^b\nabla_bu_a$ is the 4-acceleration. The rest of the variables seen on the right-hand side of (\ref{Nbua}) are the relativistic analogues of the volume scalar, the shear and the vorticity tensors. Here, these are defined by $\Theta={\rm D}^au_a$, $\sigma_{ab}={\rm D}_{\langle b}u_{b\rangle}$ and $\omega_{ab}={\rm D}_{[b}u_{a]}$ respectively.

In Einstein's theory, gravity is monitored by the associated field equations, where the matter contribution comes via its energy-momentum tensor. When dealing with a general imperfect fluid and relative to the $u_a$-field, the stress-energy tensor takes the form $T_{ab}=\rho u_au_b+ph_{ab}+2q_{(a}u_{b)}+\pi_{ab}$, where $q_a$ represents the energy flux of the matter. Substituting the latter into the (twice contracted) Bianchi identities leads to the relativistic conservation laws
\begin{equation}
\dot{\rho}= -\Theta(\rho+p)- {\rm D}^aq_a- 2A^aq_a- \sigma^{ab}\pi_{ab}  \label{edcl}
\end{equation}
and
\begin{equation}
(\rho+p)A_a= -{\rm D}_ap- h_a{}^b\dot{q}_b- {4\over3}\,\Theta q_a- (\sigma_{ab}+\omega_{ab})q^b-  {\rm D}^b\pi_{ab}- \pi_{ab}A^b\,,  \label{mdcl}
\end{equation}
of the energy and the momentum densities respectively~\cite{TCM,EMM}. A simple comparison of the above to their Newtonian counterparts, namely to Eqs.~(\ref{cont}) and (\ref{Euler}) respectively, reveals the considerable difference between the two sets of formulae. For the purposes of this work, the key difference comes from the energy-flux terms on the right-hand sides of (\ref{edcl}) and (\ref{mdcl}), none of which appears in their corresponding Newtonian counterparts. As we will see in \S~\ref{sNEtq} and \S~\ref{sREtq} later, the aforementioned flux terms, which in Eq.~(\ref{mdcl}) act as an effective (non-gravitational) force, will play the pivotal role in separating the Newtonian from the relativistic results.

\section{Peculiar velocities}\label{sPVs}
Newton's theory treats space and time as separate and absolute entities, which are the same for all observers irrespective of their motion. In relativity this is no longer the case. There, relatively moving observers ``measure'' their own space and time. In what follows, we will apply these fundamentally different approaches to cosmological peculiar motions.

\subsection{Galilean transformation}\label{ssGT}
Consider a Newtonian cosmological model and allow for two families of observers moving relative to each other. Let us also identify one  group with the (idealised/fictitious) observers following the Hubble flow and the other with those living in typical galaxies, like the Milky Way, drifting with respect to the smooth universal expansion. Hereafter, the coordinate systems associated with these two families of observers will be referred to as the Hubble frame (or the CMB frame) and the ``tilted'' frame respectively. The velocity vectors of these observers, $u_{\alpha}$ and $\tilde{u}_{\alpha}$ respectively, are related by the familiar Galilean transformation
\begin{equation}
\tilde{u}_a= u_{\alpha}+ \tilde{v}_{\alpha}\,,  \label{GT}
\end{equation}
where $\tilde{v}_a$ is the peculiar velocity of the tilted observers relative to their CMB counterparts.\footnote{Newtonian studies of peculiar flows typically use physical ($r^{\alpha}$) and comoving ($x^{\alpha}$) coordinates, with $r^{\alpha}=ax^{\alpha}$. The time-derivative of the latter gives $v_t=v_H+v_p$, where $v_t=\dot{r}^{\alpha}$, $v_H=Hr^{\alpha}$ and $v_p=a\dot{x}^{\alpha}$ are the total, the Hubble and the peculiar velocities respectively. On an FRW background, the above relation coincides with (\ref{GT}).} Since $\tilde{v}_{\alpha}$-field is the local peculiar velocity, $\tilde{u}_{\alpha}$ represents the total velocity of observers residing inside the bulk-flow domain.

The $\tilde{u}_{\alpha}$ and $u_{\alpha}$ fields also introduce a set of convective-derivative operators along their corresponding directions. In particular, following \S~\ref{ssNCA}, we define
\begin{equation}
{}^{\prime}= \partial_t+ \tilde{u}^{\alpha}\partial_{\alpha} \hspace{15mm} {\rm and} \hspace{15mm} {}^{\cdot}= \partial_t+ u^{\alpha}\partial_{\alpha}\,,  \label{NOps}
\end{equation}
as the convective derivatives in the tilted and the Hubble frames respectively.

\subsection{Lorentz boost and spacetime splitting}\label{ssLBS-TS}

\begin{figure}[tbp]
\centering \vspace{6cm} \includegraphics{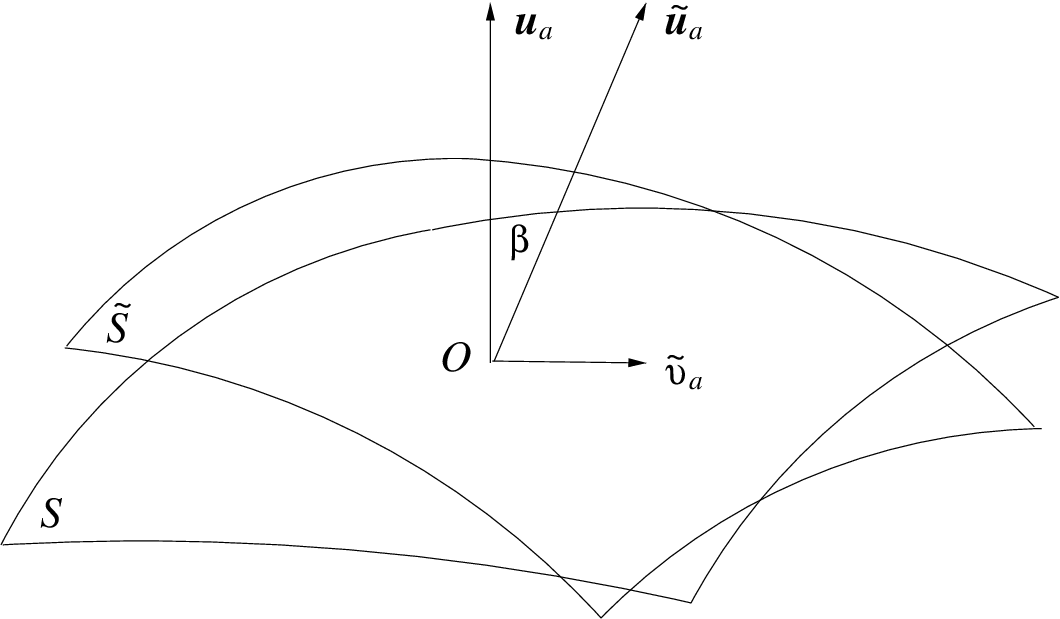} \caption{Observer ($O$) with peculiar velocity $\tilde{v}_a$ (where $\tilde{v}^2\ll1$ in our case), relative to the Hubble flow. The 4-velocities $u_a$ and $\tilde{u}_a$ (with a hyperbolic -- tilt -- angle $\beta$ between them) respectively define the CMB frame of the smooth universal expansion and that of the peculiar motion (see Eq.~(\ref{LB1})). The 3-D hypersurfaces $S$ and $\tilde{S}$ are normal to $u_a$ and $\tilde{u}_a$ and define the rest-spaces of the idealised CMB observers and of their real (tilted) counterparts respectively~\cite{T3}.}  \label{fig:bflow}
\end{figure}

In relativity the Galilean transformation is replaced by the Lorentz boost, relating the timelike 4-velocities, $\tilde{u}_a$ and $u_a$, of the relatively moving observers (see Fig.~\ref{fig:bflow}). In particular, we have
\begin{equation}
\tilde{u}_a= \tilde{\gamma}(u_a+\tilde{v}_a)\,,  \label{LB1}
\end{equation}
with $\tilde{\gamma}=(1-\tilde{v}^2)^{-1/2}$, $\tilde{v}^2=\tilde{v}_a\tilde{v}^a$ and $u_a\tilde{v}^a=0$.\footnote{The two 4-velocity fields seen in Eq.~(\ref{LB1}) form the hyperbolic angle $\beta$, with $\cosh\beta=-\tilde{u}_au^a= \tilde{\gamma}\geq1$ (see~\cite{KE} and also Fig.~\ref{fig:bflow} here). The latter determines the ``tilt'' between the two timelike directions and also explains why we use the term ``titled'' when referring to these cosmological models.} For non-relativistic peculiar motions, like those reported by all surveys, we have $\tilde{v}^2\ll1$ and $\tilde{\gamma}\simeq1$. Then, the above reduces to
\begin{equation}
\tilde{u}_a= u_a+\tilde{v}_a\,.  \label{LB2}
\end{equation}
Although the latter expression is formally identical to the Galilean transformation, there is a fundamental difference between the two formulae, since both $\tilde{u}_a$ and $u_a$ remain timelike 4-vectors, whereas their counterparts in Eq.~(\ref{GT}) are purely spatial. As discussed in \S~\ref{ssR1+3CA} earlier, the 4-velocity fields seen in (\ref{LB1}) and (\ref{LB2}) introduce an 1+3 spacetime splitting into time and 3-space, so that the corresponding observers have their own time direction (along $\tilde{u}_a$ and $u_a$ respectively) and their own 3-dimensional space ($\tilde{S}$ and $S$ orthogonal to $\tilde{u}_a$ and $u_a$ respectively -- see Fig.~\ref{fig:bflow}). The (acting) metric tensors of the aforementioned spatial hypersurfaces are the projectors $\tilde{h}_{ab}= g_{ab}+\tilde{u}_a\tilde{u}_b$ and $h_{ab}=g_{ab}+u_au_b$, with $g_{ab}$ representing the metric of the host spacetime. Then, the operator-sets
\begin{equation}
{}^{\prime}= \tilde{u}^a\nabla_a \hspace{15mm} {\rm and} \hspace{15mm} \tilde{\rm D}_a= \tilde{h}_a{}^b\nabla_b  \label{ROps1}
\end{equation}
and
\begin{equation}
{}^{\cdot}= u^a\nabla_a \hspace{15mm} {\rm and} \hspace{15mm} {\rm D}_a= h_a{}^b\nabla_b\,,  \label{ROps2}
\end{equation}
define temporal and spatial differentiation in the tilted and the Hubble frames respectively. The above spacetime splitting allows one to chose which frame to use. Here, we will perform our study in the tilted frame (see Fig.~\ref{fig:bflow} above), since it is the coordinate system of the real observers that live in a typical galaxy like ours.

\section{Linear relations between the two frames}\label{sLRBTFs}
The kinematic variables defined in the Hubble and the tilted frames are related by expressions that depend on the relative motion of the associated observers. Given that the corresponding Newtonian and relativistic relations are formally identical at the linear level, we will address them together for the economy of the presentation.

\subsection{The peculiar kinematics}\label{ssPKs}
A general velocity field can expand, or contract, it may rotate and it can also deform (i.e.~have nonzero shear). Adopting the Newtonian approach, the expansion/contraction of the three velocity fields seen in Eq.~(\ref{GT}) is monitored by the associated volume scalars, respectively defined as $\tilde{\Theta}= \partial^{\alpha}\tilde{u}_{\alpha}$, $\Theta= \partial^{\alpha}u_{\alpha}$ and $\tilde{\vartheta}= \partial^{\alpha}\tilde{v}_{\alpha}$ (see~\S~\ref{ssNCA} earlier). When using relativity, on the other hand, the associated definitions change to $\tilde{\Theta}=\tilde{\rm D}^a\tilde{u}_a$, $\Theta={\rm D}^au_a$ and $\tilde{\vartheta}=\tilde{\rm D}^a\tilde{v}_a$ (see~\S~\ref{ssR1+3CA} before). In either case, positive values imply expansion and negative ones indicate contraction. Also, in both cases, it is straightforward to show that the three scalars are related by~\cite{M}
\begin{equation}
\tilde{\Theta}= \Theta+ \tilde{\vartheta}\,.  \label{Thetas}
\end{equation}
Note that, in addition to the aforementioned differences between  the definitions, the relativistic version of the above relation is linear. This means that (\ref{Thetas}) holds for non-relativistic peculiar velocities only (see~\cite{M} for the full set of the relativistic nonlinear relations). In an expanding universe, the volume scalar $\Theta$ will always take positive values. The peculiar flow, however, can expand or contract locally, which means that $\tilde{\vartheta}$ can be either positive or negative. Nevertheless, given that that $\tilde{\vartheta}<\Theta$ in all realistic cosmological scenarios, the scalar $\tilde{\Theta}$ will be always positive and determine the expansion rate of the bulk-flow domain.

Let us now confine our study to a perturbed, almost-FRW, cosmological model filled with pressureless matter. Then, to maintain the linear nature of our analysis, we demand that $\tilde{\vartheta}/\Theta\ll1$ at all times. Following (\ref{Thetas}), this means that bulk flows with $\tilde{\vartheta}>0$ will have $\tilde{\Theta}\gtrsim\Theta$ and it will expand slightly faster than the background universe. Observers inside contracting bulk flows, on the other hand, will have $\tilde{\Theta}\lesssim\Theta$ and experience slightly slower expansion.\footnote{The volume scalar is related to the Hubble parameter. In fact, $\Theta=3H$ and $\tilde{\Theta}=3\tilde{H}$, with $H$ and $\tilde{H}$ being the Hubble parameters in the CMB and the tilted frames respectively. Then, Eq. (\ref{Thetas}) also reads $\tilde{H}=H+\tilde{\vartheta}/3$.} In either case, the agent solely responsible for these differences is the peculiar flow. Clearly, however, the relative-motion effect on the expansion is local and only affects the neighbourhood of the bulk-flow domain.

The different expansion rates between the CMB and the tilted observers seen in Eq.~(\ref{Thetas}), imply that their relative motion will induce differences in the associated deceleration/acceleration rates as well. Indeed, by differentiating (\ref{GT}) with respect to time, we obtain
\begin{equation}
\tilde{\Theta}^{\prime}= \dot{\Theta}+ \tilde{\vartheta}^{\prime}\,.  \label{Theta'}
\end{equation}
to linear order. Recall that, in the Newtonian version of the above, primes and overdots indicate convective derivatives in the ``tilted'' and the Hubble frames respectively. Put another way $\tilde{\Theta}^{\prime}=\partial_t\tilde{\Theta}+ \tilde{u}^{\alpha}\partial_{\alpha}\tilde{\Theta}$,  $\dot{\Theta}= \partial_t\Theta+u^{\alpha}\partial_{\alpha}\Theta$ and $\tilde{\vartheta}^{\prime}=\partial_t\tilde{\vartheta}+ \tilde{u}^{\alpha}\partial_{\alpha}\tilde{\vartheta}$. In relativity, on the other hand, primes and overdots indicate covariant differentiation along the corresponding 4-velocity fields. More specifically, $\tilde{\Theta}^{\prime}= \tilde{u}^a\nabla_a\tilde{\Theta}$, $\dot{\Theta}= u^a\nabla_a\Theta$ and $\tilde{\vartheta}^{\prime}= \tilde{u}^a\nabla_a\tilde{\vartheta}$. Finally, we note that the sign and the value of $\tilde{\vartheta}^{\prime}$ determine the local (volume) deceleration/acceleration of the peculiar flow itself.

\subsection{The deceleration parameters}\label{ssDPs}
The deceleration/acceleration of the expansion, as measured in the  coordinate systems of the Hubble flow and in that of the bulk peculiar motion, is monitored by the corresponding deceleration parameters. These are
\begin{equation}
\tilde{q}=- \left(1+{3\tilde{\Theta}^{\prime}\over\tilde{\Theta}^2}\right) \hspace{15mm} {\rm and} \hspace{15mm} q=- \left(1+{3\dot{\Theta}\over\Theta^2}\right)\,, \label{qs1}
\end{equation}
in the tilted and in the CMB frames respectively. The above can be used both in a Newtonian and in a relativistic study, as long as one accounts for the differences between the definitions of the volume scalars and their derivatives.  Solving (\ref{qs1}) for $\tilde{\Theta}^{\prime}$ and $\dot{\Theta}$, substituting the resulting expressions into the right-hand side of Eq.~(\ref{Theta'}) and then keeping up to $\tilde{v}$-order terms, leads to the following relation
\begin{equation}
1+ \tilde{q}= \left(1+q -{3\tilde{\vartheta}^{\prime}\over\Theta^2}\right) \left(1+{\tilde{\vartheta}\over\Theta}\right)^{-2}\,,  \label{qs2}
\end{equation}
between $\tilde{q}$ and $q$~\cite{T1,T2,T3}. Given that $\Theta=3H$ and that $\dot{H}=-H^2-\kappa\rho/6$ in the pressure-free FRW background, the above linearises to
\begin{equation}
\tilde{q}= q+ {1\over3}\left(1+{1\over2}\,\Omega\right) {\tilde{\vartheta}^{\prime}\over\dot{H}}\,,  \label{lqs}
\end{equation}
where $\Omega=\kappa\rho/3H^2$ is the density parameter. We should also note that, although $\tilde{\vartheta}/H\ll1$ throughout the linear regime, this is not necessarily true for their derivative ratio $\tilde{\vartheta}^{\prime}/\dot{H}$.

According to Eq.~(\ref{lqs}), the deceleration parameter measured by the bulk-flow observers differs from the one measured in the Hubble frame and their difference is determined by the dimensionless ratio $\tilde{\vartheta}^{\prime}/\dot{H}$ and by the density parameter of the background universe. Moreover, expression (\ref{lqs}) holds both in Newtonian and in relativistic environments, provided the differences between the corresponding definitions are observed. Of course, in the Newtonian case $\Omega=1$ always, given the Euclidean nature of the space.

\section{Newtonian effects on $\tilde{q}$}\label{sNEtq}
In Newtonian theory, gravity is a force triggered by spatial variations in the gravitational potential. The latter couples to the density of the matter via Poisson's equation. Next, we will discuss the implications of the Newtonian approach for the linear local expansion rate of the bulk-flow observers and for the related acceleration/deceleration rate.

\subsection{The role of the gravitational potential}\label{ssRGP}
The linear nature of our study demands that $\tilde{\vartheta}/H\ll1$ always, which in turn ensures that the effect of this ratio on the deceleration parameter in Eq.~(\ref{qs2}) is negligible. This leaves the peculiar volume deceleration/acceleration (i.e.~the scalar $\tilde{\vartheta}^{\prime}$) as the only realistic possibility for a  measurable relative-motion effect on $\tilde{q}$. Taking the convective derivative of $\tilde{\vartheta}$, in the tilted frame, recalling that $\tilde{\vartheta}= \partial^{\alpha}\tilde{v}_{\alpha}$ and keeping up to linear-order terms, we arrive at
\begin{eqnarray}
\tilde{\vartheta}^{\prime}= -H\tilde{\vartheta}+ \partial^{\alpha}\tilde{v}^{\prime}_{\alpha}\,,  \label{Nltvtheta'1}
\end{eqnarray}
to first approximation. The above reveals how the background expansion slows down the growth of $\tilde{\vartheta}$, by acting as (effective) friction. Also, not surprisingly, the temporal evolution of the peculiar volume scalar depends on the time-derivative of the peculiar velocity. In Newtonian theory, the latter satisfies the linear evolution formula~\cite{FT}
\begin{equation}
\tilde{v}^{\prime}_{\alpha}= -H\tilde{v}_{\alpha}- \partial_{\alpha}\Phi\,,  \label{Nltv'}
\end{equation}
which ensures that $\partial_{\alpha}\Phi$ is the sole source of linear peculiar-velocity perturbations. The above relation is identical to those obtained in typical (not necessarily covariant) studies (e.g.~see~\cite{Pe2,Netal}), provided the latter are written in physical rather than comoving coordinates.

\subsection{Newtonian relative-motion effects on 
$\tilde{q}$}\label{ssNR-MEtq}
Adopting the ``Jeans' swindle'' (e.g.~see~\cite{BT}), the (perturbed) gravitational potential couples to matter perturbations by means of the linearised Poisson equation
\begin{equation}
\partial^2\Phi= {1\over2}\,\kappa\rho\delta\,,  \label{Poisson}
\end{equation}
where $\delta=\delta\rho/\rho$ is the familiar density contrast. Substituting (\ref{Nltv'}) back into the right-hand side of Eq.~(\ref{Nltvtheta'1}) and then using (\ref{Poisson}), leads to
\begin{equation}
\tilde{\vartheta}^{\prime}= -2H\tilde{\vartheta}- {1\over2}\,\kappa\rho\delta\,.  \label{Nltvtheta'2}
\end{equation}
This result shows that perturbations in the density distribution can (in principle at least) force bulk flows to contract or expand locally. As expected, overdensities in the matter distribution (i.e.~those with $\tilde{\delta}>0$) cause local contraction, whereas underdensities lead to expansion.

Our final step is to combine Eqs.~(\ref{lqs}) and (\ref{Nltvtheta'2}). In particular, recalling that $H^2=\kappa\rho/3$ in the (Newtonian) FRW background and keeping up to first-order terms, we arrive at
\begin{equation}
\tilde{q}\simeq q+ {1\over2}\,\delta\,.  \label{Nlqs2}
\end{equation}
The above is the linear Newtonian relation between the deceleration parameter ($\tilde{q}$) measured in the rest-frame of typical observers in the universe and the deceleration parameter ($q$) measured by their idealised counterparts following the Hubble frame. Assuming that $q$ is of order unity, expression (\ref{Nlqs2}) guarantees that $\tilde{q}\simeq q$ throughout the linear regime, during which $\delta\ll1$. Therefore, within linear Newtonian cosmological perturbation theory, the relative-motion effects cannot change the (local) sign of the deceleration parameter. Put another way, the Newtonian effects cannot make a decelerating universe appear accelerating and vice-versa.\footnote{Taken at face value, relation (\ref{Nlqs2}) seems to suggest that $\tilde{q}$ could take negative values in low-density domains that expand faster than the background universe, namely in voids with $\delta\rightarrow-1$ and $\tilde{\vartheta}>0$. In that case $\tilde{q}$ can become marginally negative, even when $q\simeq1/2$. Having said that, one should be very cautious before applying linear results to nonlinear structures, like the large-scale voids. An alternative (also unlikely) possibility occurs when $0<q\ll1$, in which case $\tilde{q}$ turns negative when $\delta<0$ and $|\delta|>q$.}

\section{Relativistic effects on $\tilde{q}$}\label{sREtq}
The Newtonian relative-motion effects on the deceleration parameter come via the gravitational potential and the Poisson equation. In relativity there is no potential, but spacetime curvature. Also, Poisson's formula has been replaced by the Einstein field equations. In what follows, we will examine whether these differences can alter the Newtonian results of the last section.

\subsection{The implications of the matter's bulk 
motion}\label{ssIMBM}
Earlier, in \S~\ref{sLRBTFs}, we demonstrated that the Newtonian and the relativistic studies proceed in parallel, at least up to a certain point. More specifically, expression (\ref{lqs}) holds in both theories, provided the differences in the definitions of the volume scalars and their derivatives are accounted for. The two approaches start to diverge when the gravitational field comes into play. In Newtonian gravity, the latter propagates via the associated potential. This is no longer the case in relativity, where gravity is the manifestation of spacetime curvature. Newtonian physics also demands that the specifics of the matter component (namely its density, pressure, etc) are independent of the observer's motion. Again, this is not true in Einstein's theory, where relatively moving observers generally disagree on the nature of the matter fields involved, even when the associated peculiar velocities are small. Indeed, to linear order, the relations between the matter variables in the tilted and the Hubble frames are~\cite{M}
\begin{equation}
\tilde{\rho}= \rho\,, \hspace{10mm} \tilde{p}= p\,, \hspace{10mm} \tilde{q}_a= q_a- (\rho+p)\tilde{v}_a \hspace{10mm} {\rm and} \hspace{10mm} \tilde{\pi}_{ab}= \pi_{ab}\,.  \label{Rrels}
\end{equation}
Here, $\rho$ is the energy density, $p$ is the isotropic pressure, $q_a$ is the energy flux and $\pi_{ab}$ is the anisotropic pressure (the viscosity) of the matter in the coordinate system of the Hubble expansion. The tilded variables, on the other hand, are measured in the bulk-flow frame of the real observers. Therefore, although the density and the pressure (both isotropic and anisotropic) of the cosmic medium remain the same between the two frames, the energy flux changes. This means that a fluid that appears perfect to the Hubble observers will look imperfect to their tilted counterparts. Indeed, following (\ref{Rrels}c), we have $\tilde{q}_a= -(\rho+p)\tilde{v}_a\neq0$ when $q_a=0$.

At this point it is important to underline that, contrary to Newtonian gravity, in relativity the flux vector also contributes to the energy-momentum tensor of the matter (see \S~\ref{ssR1+3CA} earlier). In other words, the energy flux of the peculiar motion has an additional input to the relativistic gravitational field. One might therefore say that the bulk flow itself gravitates~\cite{TT,FT}. In what follows, we will discuss the implications of such a fundamental difference between Newtonian gravity and general relativity for our analysis.

\subsection{The role of the bulk-flow flux}\label{ssRB-FF}
The contribution of the ``peculiar'' energy flux ($\tilde{q}_a$) to the field equations of general relativity, as noted above, feeds into the conservation laws (see Eqs~(\ref{edcl}) and (\ref{mdcl}) in \S~\ref{ssR1+3CA}) via the (twice contracted) Bianchi identities. Then, the spatial gradient of (\ref{edcl}) leads to the relativistic propagation formula of density inhomogeneities (see Eq.~(2.3.1) in~\cite{TCM} or Eq.~(10.101) in~\cite{EMM}). Linearising either of these two propagation equations in the tilted frame, while assuming a pressureless FRW cosmology as our background model, leads to
\begin{equation}
\tilde{\Delta}_a^{\prime}= -\tilde{\mathcal{Z}}_a+ {3aH\over\rho}\left(\tilde{q}_a^{\prime}+4H\tilde{q}_a\right)- {a\over\rho}\,\tilde{\rm D}_a\tilde{\rm D}^b\tilde{q}_b\,,  \label{RltDelta'}
\end{equation}
where $\tilde{\Delta}_a=(a/\rho)\tilde{\rm D}_a\tilde{\rho}$ and $\tilde{\mathcal{Z}}_a=a\tilde{\rm D}_a\tilde{\Theta}$ respectively describe inhomogeneities in the matter density and in the universal expansion (see Eq.~(10.152) in~\cite{EMM}).

A comparison between (\ref{RltDelta'}) and its Newtonian counterpart is due here. Following~\cite{El3}, the spatial gradient of the Newtonian continuity equation (see (\ref{cont}) in \S~\ref{ssNCA}) gives
\begin{equation}
\tilde{\Delta}_{\alpha}^{\prime}= -\tilde{\mathcal{Z}}_{\alpha}\,,  \label{NltDelta'}
\end{equation}
to first order. The latter monitors the evolution of linear density inhomogeneities at the Newtonian limit and in the bulk-flow frame. Note that $\tilde{\Delta}_{\alpha}= (a/\rho)\partial_{\alpha}\rho$ and $\tilde{\mathcal{Z}}_{\alpha}= a\partial_{\alpha}\tilde{\Theta}$, while the primes denote convective derivatives in the tilted frame. The difference between Eqs.~(\ref{RltDelta'}) and (\ref{NltDelta'}) is profound and it is entirely due to the bulk-flow flux, which contributes only to the relativistic expression. Recall that the aforementioned difference reflects the one between the Newtonian and the relativistic conservation laws discussed in \S~\ref{ssNCA} and \S~\ref{ssR1+3CA} earlier. The lack of any flux terms in (\ref{NltDelta'}) is the reason Newton's and Einstein's theories arrive at so different results and conclusions, when applied to peculiar-velocity perturbations (see also~\cite{FT}).

Going back to general relativity, keeping in mind that  $\tilde{q}_a=-\rho\tilde{v}_a$ in the tilted frame (see Eq.~(\ref{Rrels}c)) and taking the spatial divergence of (\ref{RltDelta'}), we arrive at the linear relation
\begin{equation}
\tilde{\vartheta}^{\prime}= -2H\tilde{\vartheta}+ {1\over3H}\,\tilde{\rm D}^2\tilde{\vartheta}- {1\over3a^2H}\,\left(\tilde{\Delta}^{\prime} +\tilde{\mathcal{Z}}\right)\,,  \label{Rltvtheta'}
\end{equation}
with ${\rm D}^a={\rm D}^a{\rm D}_a$ being the covariant spatial Laplacian~\cite{TK}. In addition, $\tilde{\Delta}=a\tilde{\rm D}^a\tilde{\Delta}_a$ and $\tilde{\mathcal{Z}}=a\tilde{\rm D}^a\tilde{\mathcal{Z}}_a$ describe scalar inhomogeneities in the matter distribution and in the expansion respectively, with the former ($\Delta$) directly corresponding to the familiar density contrast ($\delta$) used in the Newtonian analysis (see \S~\ref{ssNR-MEtq} earlier). Again, the profound difference between (\ref{Rltvtheta'}) and its Newtonian counterpart (compare to Eq.~(\ref{Nltvtheta'2}) in \S~\ref{ssNR-MEtq}) is entirely due to the peculiar motion of the matter. More specifically, the additional terms on the right-hand side of (\ref{Rltvtheta'}) can be traced back to the purely relativistic flux contribution to the energy-momentum conservation laws (see \S~\ref{ssIMBM} before), as well as to expression (\ref{RltDelta'}) above.

\subsection{Relativistic relative-motion effects on 
$\tilde{q}$}\label{ssRR-MEq}
Given the clear difference between (\ref{Rltvtheta'}) and its Newtonian analogue, namely Eq.~(\ref{Nltvtheta'2}) in \S~\ref{ssNR-MEtq}, we expect the relativistic effects on $\tilde{q}$ to differ as well. Indeed, substituting (\ref{Rltvtheta'}) into Eq.~(\ref{lqs}) and taking into account that $\tilde{\vartheta}/H\ll1$ at the linear level, the latter formula recasts to
\begin{equation}
\tilde{q}= q+ {2\over3H}\,\tilde{\vartheta}- {1\over9H^3}\,\tilde{\rm D}^2\tilde{\vartheta}+ {1\over9a^2H^3} \left(\tilde{\Delta}^{\prime}+\tilde{\mathcal{Z}}\right)\,.  \label{Rlqs2}
\end{equation}
This linear expression provides the deceleration parameter measured by observers living in typical galaxies, like our Milky way, which move relative to the Hubble expansion of a perturbed almost-FRW universe filled with pressureless dust. The difference between (\ref{Rlqs2}) and the corresponding Newtonian relation (compare to Eq.~(\ref{Nlqs2}) in \S~\ref{ssNR-MEtq}) is more than obvious. Moreover, even within the linear regime, the relativistic correction can be strong because of the Laplacian term on the right-hand side of (\ref{Rlqs2}). Indeed, a simple harmonic decomposition of the latter leads to the following expression
\begin{equation}
\tilde{q}_{(n)}= q+ {2\over3}\,\left[1+ {1\over6}\left({\lambda_H\over\lambda_{(n)}}\right)^2\right]\, {\tilde{\vartheta}_{(n)}\over H}+ {1\over9}\left({\lambda_H\over\lambda_K}\right)^2 \left({\tilde{\Delta}_{(n)}^{\prime}\over H} +{\tilde{\mathcal{Z}}_{(n)}\over H}\right)\,,  \label{Rlhqs1}
\end{equation}
for the $n$-th harmonic.\footnote{We have employed the familiar harmonic decomposition for the perturbations (e.g.~$\tilde{\vartheta}= \sum_n\tilde{\vartheta}_{(n)}\mathcal{Q}_{(n)}$), where $\tilde{\rm D}_a\tilde{\vartheta}_{(n)}=0=\mathcal{Q}^{\prime}_{(n)}$ and $\tilde{\rm D}^2\mathcal{Q}_{(n)}=-(n/a)^2\mathcal{Q}_{(n)}$ by construction. Note that $n$ represents the comoving wavenumber of the harmonic mode (with $n>0$), which makes $\lambda_{(n)}=a/n$ its physical wavelength.} In the above $\lambda_{(n)}=a/n$ is the physical scale of the peculiar-velocity perturbation, $\lambda_H= 1/H=3/\Theta$ is the Hubble horizon and $\lambda_K=a/|K|$ is the curvature scale of the universe (with $K=\pm1$ being the associated 3-curvature index). In a nearly flat FRW universe, with $\Omega\rightarrow1$, this translates into $(\lambda_H/\lambda_K)^2= |1-\Omega|\rightarrow0$. On these grounds, the last term on the right-hand side of (\ref{Rlhqs1}) is negligible, leaving
\begin{equation}
\tilde{q}_{(n)}= q+ {1\over9}\left({\lambda_H\over\lambda_{(n)}}\right)^2\, {\tilde{\vartheta}_{(n)}\over H}\,.  \label{Rlhqs2}
\end{equation}
This is the deceleration parameter measured by observers living inside a bulk flow, with physical size $\lambda_{(n)}$ and moving relative to the Hubble expansion of a perturbed Einstein-de Sitter universe~\cite{TK,T3}. The comparison with the corresponding Newtonian relation (see Eq.~(\ref{Nlqs2}) in \S~\ref{ssNR-MEtq}), immediately shows that the relativistic effect can be large. Indeed, even though $\tilde{\vartheta}_{(n)}/H\ll1$ throughout the linear regime, the effect of the correction term on the right-hand side of (\ref{Rlhqs2}) grows increasingly stronger as we move to progressively smaller scales (where $\lambda_H/\lambda_{(n)}\gg1$). Moreover, in qualitative terms, the effect depends on the sign of $\tilde{\vartheta}_{(n)}$ as well. When the latter is positive we have $\tilde{q}_{(n)}>q$, while in the opposite case $\tilde{q}_{(n)}<q$. This means that in slightly expanding bulk flows (i.e.~those with $0<\tilde{\vartheta}_{(n)}/H\ll1$) the deceleration parameter measured in the tilted frame is larger than the one measured in the Hubble frame. On the other hand, observers inside slightly contracting bulk flows (with $-1\ll\tilde{\vartheta}_{(n)}/H<0$), will measure smaller deceleration parameter than their Hubble-flow counterparts.

Following (\ref{Rlhqs2}), the relative motion effects dominate its right-hand side when the ``correction term'' equals (in absolute value) the deceleration parameter measured in the Hubble frame. This happens at a characteristic scale given by~\cite{T3}
\begin{equation}
\lambda_{(n)}\equiv \lambda_P= \sqrt{{1\over9q}{|\tilde{\vartheta}_{(n)}|\over H}}\, \lambda_H\,,  \label{lambdaP}
\end{equation}
since $q>0$ always in Friedmann universes with conventional matter. The above length defines the domain inside which the linear kinematics are determined by peculiar-velocity perturbations rather than by the background universal expansion. In this respect, $\lambda_P$ closely resembles the familiar Jeans length ($\lambda_J$), marking the scale below which pressure-gradient perturbations dominate over the background gravity and thus dictate the linear evolution of density inhomogeneities. We will therefore refer to $\lambda_P$ as the ``peculiar Jeans length'' to underline the aforementioned close analogy (see~\cite{T3} for further details and discussion).

\begin{figure}[tbp]
\centering \vspace{4cm} \includegraphics{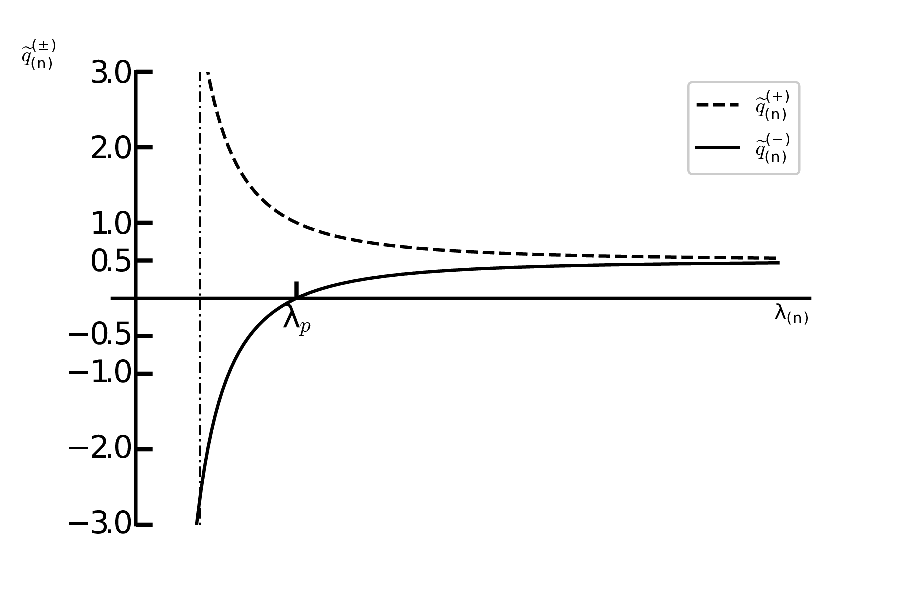} \caption{The scale-distribution of the deceleration parameter ($\tilde{q}^{(\pm)}_{(n)}$) in the rest-frame of a bulk flow with peculiar Jeans length $\lambda_P$ (see Eq.~(\ref{Rlhqs3})). On large scales, where $\lambda_{(n)}\gg\lambda_P$, the local deceleration parameter approaches that of the background Einstein-de Sitter universe (i.e.~$\tilde{q}_{(n)}^{(\pm)}\rightarrow q=0.5$), but n smaller scales it diverges. In (slightly) expanding peculiar flows (dashed curve), $\tilde{q}_{(n)}^{(+)}>1/2$, with the deceleration parameter crossing the $\tilde{q}_{(n)}^{(+)}=1$ mark at $\lambda_{(n)}=\lambda_P$. Inside contracting bulk peculiar motions (solid curve), on the other hand, $\tilde{q}_{(n)}^{(-)}<1/2$. There, the local deceleration parameter drops below the $\tilde{q}_{(n)}^{(-)}=0$ threshold at the transition scale, namely at $\lambda_P$. The dotted vertical line indicates the nonlinear cutoff scale, below which our linear analysis no longer holds. The nonlinear threshold is estimated below the 100~Mpc mark, which is typically much smaller than $\lambda_P$ (see Eqs.~(\ref{lambdaP1}) and (\ref{lambdaP2}) in \S~\ref{ssETS} next). Note that we have adopted the $\lambda_P=1$ normalisation.}  \label{fig:tq+-}
\end{figure}

An alternative expression is obtained after combing (\ref{Rlhqs2}) with definition (\ref{lambdaP}) More specifically, substituting the latter into the former leads to the following compact relation
\begin{equation}
\tilde{q}^{(\pm)}_{(n)}= q\left[1\pm\left({\lambda_P\over\lambda_{(n)}}\right)^2\right]\,,  \label{Rlhqs3}
\end{equation}
between the two deceleration parameters (see also Fig.~\ref{fig:tq+-} for the related diagram). Recall that the $+/-$ signs corresponds to expanding/contracting bulk flows (i.e.~those with $\tilde{\vartheta}_{(n)}\gtrless0$ respectively). Consequently, on scales much larger than the peculiar Jeans length, where $\lambda_P/\lambda_{(n)}\ll1$, the correction due to the observers relative motion is negligible and the two deceleration parameters essentially coincide (i.e.~$\tilde{q}^{(\pm)}_{(n)}\rightarrow q$). This agrees with our expectation that the peculiar-motion effects fade away as we move on to progressively larger lengths. On scales smaller than $\lambda_P$, however, the situation is quite different. There, observers inside (slightly) expanding bulk motions will measure $\tilde{q}_{(n)}>2q$. In (slightly) contracting peculiar flows, on the other hand, the associated observers will measure $\tilde{q}_{(n)}<0$. In this latter case, the peculiar Jeans length also marks the ``transition scale'', where the deceleration parameter turns from positive to negative (see Fig.~\ref{fig:tq+-} as well as~\cite{T3}). Although the numerics (e.g.~the location of the transition length) depend on the specifics of the bulk flow in measurements (namely on the local expansion/contraction rate ($\tilde{\vartheta}_{(n)}$)), the profile of the solid curve depicted in Fig.~\ref{fig:tq+-} is a characteristic (signature-like) feature of the distribution of $\tilde{q}^{(-)}_{(n)}$ in the bulk-flow scenario~\cite{T3}. More specifically, the evolution of $\tilde{q}^{(-)}_{(n)}$ in terms of scale is nonlinear. In fact, the deceleration parameter remains nearly constant (with $\tilde{q}^{(-)}_{(n)}\simeq q=0.5$) on large scales (far away from the observer) and becomes increasingly more negative on small scales (close to the observer). Also note that the shape of the solid curve seen in Fig.~\ref{fig:tq+-} closely resembles the profile of the deceleration parameters reconstructed from the SN~Ia data (e.g.~see~\cite{GW}-\cite{AMD}). There, however, the authors introduced a (typically) two-parameter ansatz for $q$. Here, the profile of the $\tilde{q}$-distribution follows naturally from the relative motion effects, as encoded in Eq.~(\ref{Rlhqs3}).

\subsection{Estimating the transition scale}\label{ssETS}
Expressions (\ref{Rlhqs2}) and (\ref{lambdaP}) provide the local value of the deceleration parameter measured inside a given bulk flow and the corresponding peculiar Jeans length. Both depend on the local expansion/contraction rate ($\tilde{\vartheta}$ -- hereafter we will drop the mode-index ($n$)) of the peculiar motion. However, observations provide the mean bulk velocity ($\langle\tilde{v}\rangle$) of the drift flow and not its spatial divergence ($\tilde{\vartheta}$), primarily because of noise-related problems. Nevertheless, using standard dimensional-analysis arguments, we may set $\tilde{\vartheta}\sim \langle\tilde{v}\rangle/\lambda$ and estimate $\tilde{\vartheta}$ (in km/secMpc) from the reported mean velocities and scales ($\langle\tilde{v}\rangle$ and $\lambda$ respectively). Then, once the value of the Hubble parameter is given, we can use data from bulk-flow surveys to estimate $\tilde{q}^{(\pm)}$ and $\lambda_P$. The reader is referred to Table~1 in~\cite{T3} for further discussion and for a list of representative estimates based on bulk-flow measurements. Typically, these report peculiar velocities of few hundred km/sec on scales around hundred Mpc (e.g.~see~\cite{CMSS}-\cite{SYFB}), with the (estimated) dimensionless ratio $\tilde{\vartheta}/H$ varying between 0.01 and 0.1 (for $H\simeq70$~km/secMpc). Then, assuming an Einstein-de Sitter background (with $q=0.5$ everywhere), definition (\ref{lambdaP}) gives
\begin{equation}
\lambda_P\simeq 190~{\rm Mpc}\,, \hspace{5mm} {\rm when} \hspace{5mm} \tilde{\vartheta}/H\simeq 0.01\,, \hspace{10mm} \lambda_P\simeq 420~{\rm Mpc}\,, \hspace{5mm} {\rm when} \hspace{5mm} \tilde{\vartheta}/H\simeq 0.05  \label{lambdaP1}
\end{equation}
and
\begin{equation}
\lambda_P\simeq 600~{\rm Mpc}\,, \hspace{5mm} {\rm when} \hspace{5mm} \tilde{\vartheta}/H\simeq 0.1\,. \label{lambdaP2}
\end{equation}
In every case, $\lambda_P$ marks the scale below which the relative motion effects dominate over the background Hubble expansion and determine the local kinematics.\footnote{Following definition (\ref{lambdaP}), once the background values of $H$ and $q$ are fixed, the peculiar Jeans length ($\lambda_P$) associated with any given bulk flow depends on the latter's local expansion/contraction rate ($\tilde{\vartheta}$). Also note that, the lower the Hubble parameter, the stronger the relative-motion effects and the larger the value of $\lambda_P$.} When dealing with locally expanding bulk flows, $\lambda_P$ defines the $\tilde{q}^{(+)}=1$ threshold (dashed curve in Fig.~\ref{fig:tq+-}). For locally contracting bulk motions, on the other hand, $\lambda_P$ sets the ``transition scale''. There, as seen by an observer located at the centre of the bulk-flow domain, $\tilde{q}^{(-)}$ turns from positive to negative (solid curve in Fig.~\ref{fig:tq+-}).

\begin{figure}[tbp]
\centering \vspace{3.5cm} \includegraphics{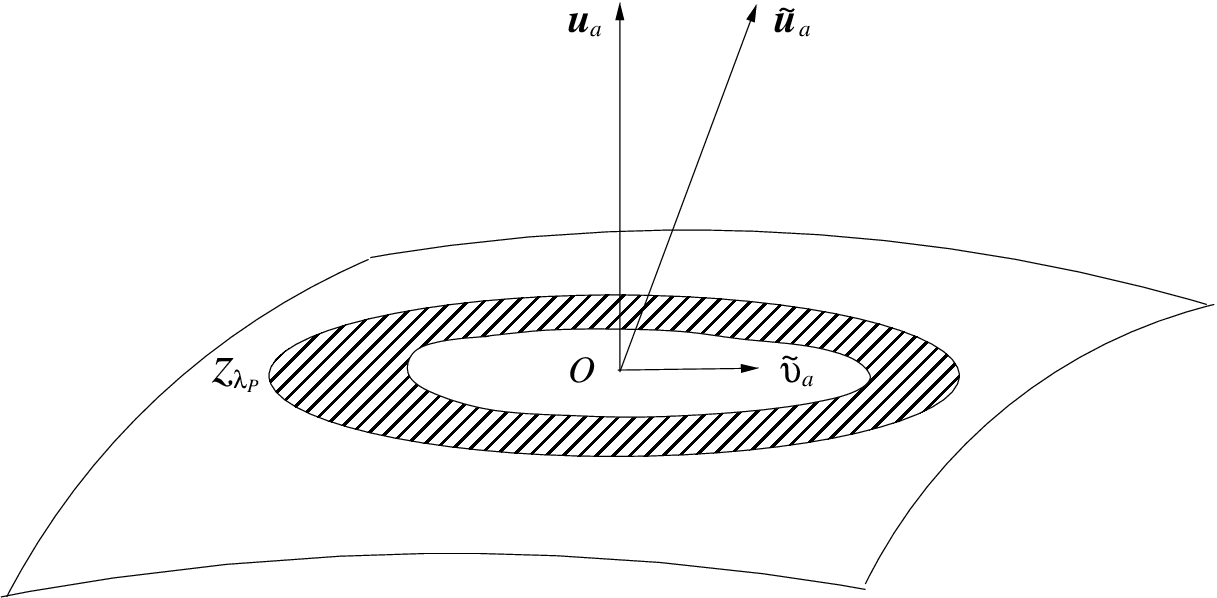} \caption{Observer ($O$) with 4-velocity $\tilde{u}_a$, at the centre of a bulk flow (central region), like those reported in~\cite{CMSS}-\cite{SYFB}, moving with peculiar velocity $\tilde{v}_a$, relative to the smooth Hubble expansion (identified with the $u_a$-field) of an Einstein-de Sitter universe with $q=0.5$. The shaded spherical region is determined by the peculiar Jeans length ($\lambda_P$). The latter corresponds to redshift $z_{\lambda_P}$, beyond which the deceleration parameter is always positive. In the case of contracting bulk flows, $z_{\lambda_P}$ also marks the transition redshift where the sign of the deceleration parameter ($\tilde{q}^{(-)}$) changes from positive to negative (see also Fig.~\ref{fig:tq+-}). From the observer's point of view, $\tilde{q}^{(-)}$ appears to become progressively less negative with increasing redshift, it crosses the zero-mark point at $z_{\lambda_P}$, turns positive beyond that threshold and eventually approaches $\tilde{q}^{(-)}\rightarrow q=0.5$ at large enough redshifts (see Fig.~\ref{fig:tq+-}). An unsuspecting bulk-flow observer may then be mislead to believe that their universe started accelerating at $z=z_{\lambda_P}$ and in so doing misinterpret the local change in the sign of $\tilde{q}^{(-)}$ as the onset of recent global acceleration~\cite{T3}.}   \label{fig:PJeans}
\end{figure}

The negative values of $\tilde{q}^{(-)}$ (on scales smaller than the associated transition length) are mere local measurements, triggered by relative-motion effects. Globally, the universe is still decelerating with $q=0.5$. Therefore, the acceleration ``experienced'' by the associated bulk-flow observers is a mere artifact of their peculiar motion relative to the smooth Hubble flow.\footnote{The same is also true for the apparent over-deceleration (with $\tilde{q}^{(+)}>1$) measured on scales smaller than $\lambda_P$ by observers inside locally expanding bulk flows. Globally, the universe is decelerating with $q=0.5$.} Nevertheless, the affected scales are large enough ($\lambda_P$ typically varies between few and several hundred Mpc -- see Eqs.~(\ref{lambdaP1}) and (\ref{lambdaP2})), to make this local effect appear as a recent global event (see Fig.~\ref{fig:PJeans} and also~\cite{T3} for further comments).

It should be noted that we have set $H\simeq70$~km/secMpc today in (\ref{lambdaP1}) and (\ref{lambdaP2}) purely for demonstration purposes. One should keep in mind, however, this value for the Hubble parameter in an Einstein-de Sitter cosmology implies an age for the universe smaller than the one typically attributed to the oldest globular clusters. The age problem appears to be alleviated in the $\Lambda$CDM scenario, though not entirely (e.g.~\cite{PS}). Nevertheless, there are still uncertainties related to both measurements. For instance, there are systematic errors in estimating the age of the oldest globular clusters, primarily due to theoretical stellar-modelling uncertainties (e.g.~\cite{Vetal}). The current value of the Hubble parameter has also been under scrutiny recently. Most of the reported measurements vary around 70~km/secMpc, but there are also estimates giving considerably lower values (see~\cite{PS} for a comprehensive list). More refined measurements of the Hubble parameter should help to resolve the current tension, which is key to the viability of essentially all cosmological models. For our purposes, in addition to the important age issue, the value of Hubble parameter also determines the strength of the relative-motion effects.\footnote{The considerable differences between the Newtonian and the relativistic treatments of peculiar velocities, as well as the underlying theoretical reasons responsible for them, hold irrespective of the current Hubble value and the age of the universe.} In particular, the smaller the current value of $H$, the larger the $\tilde{\vartheta}/H$-ratio and therefore the stronger the aforementioned effects.

\subsection{Characteristic features}\label{ssCFs}
One may wonder whether there are characteristic imprints of the relative-motion effects discussed in this study that one should look for in the data. According to our analysis, the first (qualitative) sign is in the scale-distribution of $\tilde{q}^{(-)}$, which should generally follow the profile of the solid curve depicted in Fig.~\ref{fig:tq+-}. The latter argues for a nonlinear scale-evolution of the deceleration parameter, with $\tilde{q}^{(-)}$ remaining essentially constant far away from the observer (assumed to be near the centre of the bulk flow) and becoming increasingly more negative in their close vicinity. Such a behaviour agrees with our expectation that the impact of peculiar motions should be strong in the observers local neighbourhood and it should fade away on progressively larger distances. Moreover, the profile of the solid curve depicted in Fig.~\ref{fig:tq+-} is in qualitative agreement with those of the deceleration parameters  reconstructed from the supernovae data~\cite{GW}-\cite{AMD}.

There is an additional characteristic imprint that is also the``trademark'' signature of relative motion, namely an apparent dipolar anisotropy in the sky-distribution of $\tilde{q}^{(-)}$, triggered by the observers' peculiar flow. For instance, the dipole seen in the CMB spectrum is believed to be such a Doppler-like effect due to the drift motion of our Milky Way (and of the Local Group) relative to the smooth Hubble expansion. A similar dipolar anisotropy should appear in the measured distribution of $\tilde{q}^{(-)}$, if the recent acceleration of the universe is just a local illusion caused by our peculiar motion. More specifically, the deceleration parameter should take more negative values towards a given direction in the sky and equally less negative in the opposite. Put another way, the universe should appear to accelerate faster in one direction and equally slower in the opposite. In addition, the axis of the $\tilde{q}^{(-)}$-dipole should not lie very far from that of its CMB counterpart, if they are both artifacts of our peculiar motion~\cite{T2}. Over the last decade, there have been reports that such a dipolar anisotropy (a preferred axis close to the CMB dipole) might actually exist in the data~\cite{SW}-\cite{BBA}, though none of the aforementioned works made any direct association with relative-motion effects. This connection was made recently in~\cite{CMRS1}, where a fairly strong dipole in the sky-distribution of the deceleration parameter was reported. The dipolar axis was also closely aligned with that in the CMB, while at the same time the statistical significance of the $q$-monopole dropped, thus increasing the chances the inferred universal acceleration to be a local artefact of our peculiar motion. Future observations and better data should provide an answer to the $q$-dipole debate~\cite{RH,CMRS2}. At this point, we must note that a similar dipolar anisotropy, of potentially kinematic origin too, was recently reported in the sky-distribution of the Hubble parameter as well~\cite{Metal1,Metal2}. These latter findings could also provide support to the bulk-flow scenario, since a dipolar anisotropy in the Hubble parameter should imply the same for the decelaration parameter (and vice versa), given the close relation between the two.

The theoretical possibility that local relative-motion effects may lead to the illusion of apparent global acceleration was originally raised in~\cite{T1,T2} and subsequently refined in~\cite{TK,T3}. Additional independent theoretical backing was recently given in the alternative approach of~\cite{H}. There, by considering a congruence of null geodesics in a general spacetime and analysing the associated luminosity distance between emitters and observers, the author concluded that local effects could create the false impression of accelerated expansion by changing the sign of the (effective) deceleration parameter~\cite{H}. There, it was also claimed that, when the luminosity distance is involved and peculiar velocities are allowed, the relative-motion effects are stronger than the purely kinematic ones identified here. In such a case, the transition scale ($\lambda_P$) will exceed the values estimated in Eqs.~(\ref{lambdaP1}) and (\ref{lambdaP2}) here. Finally, the same study also confirmed that an apparent dipole imprint should appear in the locally measured sky-distribution of the deceleration parameter, due to the observers' peculiar motion alone~\cite{H}. The reader is also referred to~\cite{MH} for numerical simulations based on that work.

\section{Discussion}\label{sD}
In contrast to Newtonian physics, relativity does not treat space and time as absolute and separate entities. In Einstein's theory, observers moving with respect to each other measure their own space, have their own time and generally experience different versions of what we may call reality. For instance, the apparent nature of the cosmic fluid changes between relatively moving observers. Those following the smooth Hubble flow may ``see'' a perfect cosmic medium, while their drifting counterparts will ``see'' it as imperfect due to relative-motion effects alone. An additional fundamental difference between the two theories is their treatment of the gravitational field. Contrary to Newtonian physics, where only the density of the matter gravitates (through Poisson's formula), general relativity allows for additional contributions, coming from the pressure (isotropic as well as anisotropic) and from the energy flux (all contributing to the energy-momentum tensor). In the case of large-scale peculiar motions, one could interpret the extra gravitational input of their energy flux by saying that, for all practical purposes, bulk flows gravitate~\cite{TT,FT}. This difference in the treatment of the gravitational field, changes significantly the relativistic conservation laws of both the energy and the momentum densities (i.e.~the continuity and the Euler equations) and eventually alters the relativistic formulae monitoring the evolution and the implications of large-scale peculiar motions.

The history of astronomy contains many examples where observations were seriously misinterpreted due to relative-motion effects. Given that no real observer in the universe follows the smooth Hubble expansion, we have considered the implications of the observers' relative motion for their measurement of the deceleration parameter. In particular, we compared the deceleration parameter measured in the Hubble/CMB frame, which coincides with that of the universe itself, with the deceleration parameter measured in the tilted frame of the bulk-flow observers. The comparison was done by working in parallel the Newtonian and the relativistic treatments, assuming a perturbed, spatially flat, Friedmann universe filled with low-energy dust. In both studies, we used closely analogous mathematical tools, namely the relativistic 1+3 covariant formalism and its Newtonian version. Our analysis showed that the two studies begin to diverge once the earlier mentioned relativistic energy-flux contribution to the local gravitational field is accounted for, and starts to modify the equations monitoring the peculiar velocity field and its impact. As a result, the final conclusions of the two approaches differ considerably, even in the linear regime.

We found that the Newtonian effects of the observers' motion relative to the smooth Hubble flow, have no significant impact on the local measurements of the deceleration parameter, at least at the linear perturbative level. The deceleration parameter measured in the bulk-flow frame is essentially identical to the one measured in the rest-frame of the mean universal expansion. The relativistic effects, on the other hand, are considerably stronger. The extra flux contribution to the stress-energy tensor leads to an effective force in the momentum conservation law, which in turn triggers scale-dependent effects on the deceleration parameter measured in the bulk-flow frame. These resemble the effects of pressure gradients (which are zero by default in our case) on density inhomogeneities and lead to a characteristic length-scale that is closely analogous to the familiar Jeans length. More specifically, this ``peculiar Jeans length'' marks the threshold below which peculiar velocity perturbations dominate over the background expansion and thus dictate the linear kinematics of the bulk-flow observers.\footnote{Recall that the standard Jeans length marks the scale below which pressure-gradient perturbations dominate over the background gravitational pull and determine the linear evolution of density inhomogeneities.} As a result, on scales smaller than $\lambda_P$, the deceleration parameter measured inside slightly expanding bulk flows can be twice as large as the deceleration parameter of the universe, or even larger. Inside slightly contracting bulk flows, on the other hand, the deceleration parameter can become negative, while the universe is still globally decelerating. Put another way, within the realm of general relativity, large-scale relative-motion effects can change the sign of the deceleration parameter. The effect is local, of course, but the affected scales are typically large enough (between few and several hundred Mpc -- see Eqs.~(\ref{lambdaP1}), (\ref{lambdaP2}) in \S~\ref{ssETS}) to create the false impression that the whole universe has recently entered an accelerated phase. Then, assuming that locally expanding and contracting bulk flows are more or less equally distributed, roughly half of the observers in the universe may think that their cosmos is over-decelerated, while the rest my be misled to believe that the expansion is under-decelerated, or even accelerated in some cases.

The signs that the recent universal acceleration may be a local artefact of our motion relative to the Hubble flow should be sought in the data. These should contain the signatures of the relative motion effects discussed in the bulk-flow scenario~\cite{T3}. The first is in the scale-distribution of the locally measured deceleration parameter, which should generally follow the nonlinear profile of the solid curve depicted in Fig.~\ref{fig:tq+-}. Similar profiles have been reconstructed from the supernovae data (e.g.~see~\cite{GW}-\cite{AMD}), though typically by introducing an ansatz for the evolution of the deceleration parameter. The second prediction of the bulk-flow scenario is the ``trademark signature'' of relative motion, namely an apparent (Doppler-like) dipolar anisotropy in the sky-distribution of the deceleration parameter. To the bulk-flow observers, the universe should appear to accelerate faster along one direction in the sky and equally slower in the opposite. Moreover, the axis of the $\tilde{q}$-dipole should lie fairly close the that of the CMB, since they are both triggered by the observers' relative motion (see~\cite{T1,T2} for the original theoretical arguments). On theoretical grounds, the presence of a dipolar anisotropy in the sky-distribution of the deceleration parameter, due to the observers' peculiar motion, was independently confirmed in the recent study of~\cite{H}. Observationally, the possible existence of a dipole axis in the universal acceleration has been argued by a number of authors over the last decade (e.g.~see~\cite{SW}-\cite{BBA}), though without directly attributing it to peculiar velocity fields. More recently, the presence of a dipole in the distribution of the deceleration parameter, triggered by relative-motion effects, was claimed in~\cite{CMRS1}. Observational support may also come from recent surveys reporting a dipolar anisotropy, of possible kinematic origin, in the sky-distribution of the Hubble parameter as well~\cite{Metal1,Metal2}.

It is therefore clear that the relativistic bulk-flow scenario requires more and further refined observational data. In addition to those mentioned above, future peculiar-velocity surveys should provide more direct estimates for the peculiar volume scalar ($\tilde{\vartheta}$). These, together with refined measurements of the Hubble parameter, will allow us to evaluate the ratio $\tilde{\vartheta}/H$, which determines the strength of the relative-motion effects. It should also be noted that a consensus on the current value of $H$ could also resolve another important issue, namely the tension between the age of the universe and that of the oldest globular clusters. The latter is key to the viability of  all cosmological scenarios, including the one outlined here (see \S~\ref{ssETS} for related comments).

Before closing our discussion, we should point out that, apart from demonstrating the relative-motion effects on the deceleration parameter and its local interpretation, the aim of this work was also to draw attention to the study of the large-scale bulk peculiar velocity fields. The latter have been largely bypassed in most theoretical treatments, or they have been investigated within the Newtonian limits. In addition, the available studies are typically performed in the idealised CMB-frame of the fictitious Hubble-flow observers. However, we have seen here that the Newtonian and the relativistic pictures can vary drastically, even at the linear perturbative level (see~\cite{FT} as well). Moreover, abandoning the coordinate system of the Hubble expansion for the tilted frame of the real observers (namely of those living in typical galaxies like the Milky Way), makes it easier to identify the relativistic effects (see also~\cite{FT}). Overall, bringing relativity into play and changing our frame of reference could also change our views on the evolution, the role and the implications of the large-scale peculiar-velocity fields.\\

\textbf{Acknowledgements:} We would like to thank Kostas Migkas, Roya Mohayaee, Mohamed Rameez and Subir Sarkar for helpful discussions and comments. This work was supported by the Hellenic Foundation for Research and Innovation (H.F.R.I.), under the ``First Call for H.F.R.I. Research Projects to support Faculty members and Researchers and the procurement of high-cost research equipment grant'' (Project Number: 789).

\end{document}